# Disruptions in the U.S. Airport Network

Steven Cao
Rutgers University
New Brunswick, NJ
scao@eden.rutgers.edu

Edoardo Conti
Rutgers University
New Brunswick, NJ
econti@eden.rutgers.edu

Alfred Thomas
Rutgers University
New Brunswick, NJ
alfredt@eden.rutgers.edu

## ABSTRACT

Our project analyzes the United States domestic airport network. We attempt to determine which airports are most vital in maintaining the underlying infrastructure for all domestic flights within the United States. To perform our analysis, we use data from the first quarter of 2010 and use several methods and algorithms that are frequently used in network science. Using these statistics, we identified the most important airports in the United States and investigate the role and significance that these airports play in maintaining the structure of the entire domestic airport network. Some of these airports include Denver International and Ted Stevens Anchorage International. We also identified any structural holes and suggested improvements that can be made to the network. Finally, through our analysis, we developed a disaster response algorithm that calculates flight path reroutes in emergency situations.

## Categories and Subject Descriptors

E.1 [*Data Structures*]: Graphs and networks; G.2.2 [*Graph Theory*]: Network problems; G.4 [*Mathematical Software*]: Algorithm design and analysis

## 1. INTRODUCTION

In this paper, we will first introduce background information and statistics on the United States airport network since the start of 2000. We will then examine where the industry stands today and discuss some common misconceptions. Although there has been some research done on this topic, the scope of previous works has largely been focused on simply analyzing the network. Our project attempts to build off of this initial analysis and suggest improvements to the underlying backbone of the network by simulating disruptions and examining the resulting impact. We also develop a novel algorithm that calculates the optimal airports that a flight should be rerouted to.

## 2. BACKRGOUND INFORMATION

After the September 11 attacks, the United States airline industry lost about 8% of its passengers over the course of the next year. To improve national security, the government enforced new security measures and regulations. However, in this paper, we will identify some of the problems that were overlooked and some of the inefficiencies in the airport network that remain today.

Of more than the 14,000 airports in the U.S., only 400 have regularly scheduled flights [1]. There are more than 1.4 billion passengers that travel through air on an annual basis in North America. However, more than 85% of all passengers traveled through only fifty airports in 2010 [2]. This relationship is illustrated below in Figure 1 and shows that passengers heavily rely on certain airports.

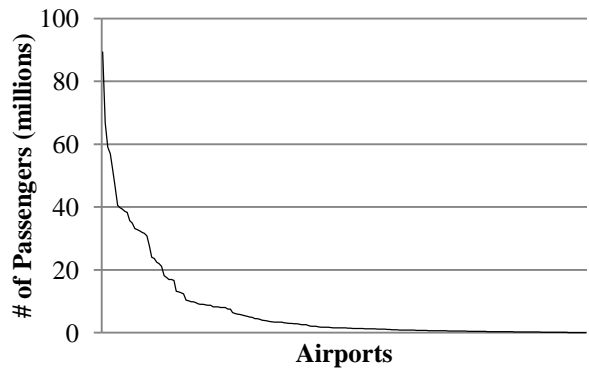

**Figure 1**

If one or more of these airports were to shut down, it would have a catastrophic impact on domestic air travel. Airports may shut down because of terrorist attacks, damages from severe weather, worker strikes, etc. From a survey that we conducted, most people believed that damages to airports in large cities such as JFK (New York), Dulles (Washington D.C.), Los Angeles International, O'Hare (Chicago), and Dallas/Fortworth would cause the most havoc to the airport network. In our analysis, we used hub, authority, PageRank and centrality measures to investigate whether this was indeed true.

## 3. RELATED WORK

There has been moderate research done on airport networks. In one related work, "Why Anchorage is not (that) important: Binary ties and Sample selection", Tore Opshal uses centrality measures such as betweeness to determine the importance of airports in international travel [3]. Other studies have also investigated the features of both international and domestic airline networks, but none have used this analysis to identify the consequences of large scale disruptions to these networks.

## 4. DATA SET

Our data was obtained from the U.S. Bureau of Transportation Statistics. From just the first quarter of 2010, there were over 2,610 nodes and 64,204 edges[1]. In our directed network, each node represents an airport and each edge represents a flight between two airports. Edge attributes include the origin of the flight, the destination of the flight, the number of passengers, and the physical distance between the origin and destination airport. Figure 2 shows a visual of the network as constructed in Cytoscape.

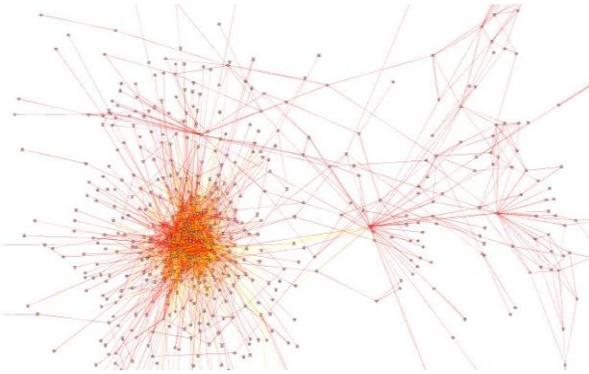

**Figure 2**

## 5. PROPOSED APPROACH

Our approach is divided up into three distinct steps. The first step of our approach is focused on the selection of the most vital airports to the U.S. domestic flight network. Using network measures such as centrality, hub score, authority score, PageRank, and articulation point detection, we compile our list of the most important U.S. domestic airports.

The second step of our approach begins by simulating disruptions in the network, specifically by disrupting the vital airports which we determined in step one. After simulating the disruption, we analyze the resulting effects on the network, particularly by noting changes in metrics such as diameter, average path length, number of strongly connected components, and number of articulation points. We aim to display the significance of disruption among our selected "important" airports and offer a comparison in disruption severity against the airports that many believe would cause the most significant network disruptions.

The third and final step of our approach involves the development and examination of a proposed solution to remediate the structural holes present in the U.S. domestic airport network. In this step, we develop a novel rerouting algorithm that takes in an airport as input and determines an efficient rerouting based on vicinity and capacity of neighboring airports. The algorithm returns rerouting options with percentages corresponding to the percent of air traffic rerouted to each determined airport. We will now examine each step in detail.

---
[1] Data collected from http://www.transtats.bts.gov/. Only first quarter was used because annual dataset was too large for our program to handle.

## 6. SELECTION OF THE MOST IMPORTANT AIRPORTS

Our selection process of the most important U.S domestic airports uses a combination of PageRank, hub scores, authority scores, centrality measures, and articulation point analysis. We begin by calculating PageRank for all nodes in the network. PageRank, in the context of the airport network, reveals the importance of a particular airport based on the probability that a sequence of random flights will allow a passenger to arrive at the airport. The airports with the top 5 PageRanks are shown in Table 1.

**Table 1**

| Top 5 PageRank Score | | |
|---|---|---|
| Rank | Airport | Score |
| 1 | Denver International | 0.0174 |
| 2 | Ted Stevens Anchorage International | 0.0169 |
| 3 | Chicago O'Hare International | 0.0151 |
| 4 | Hartsfield-Jackson Atlanta International | 0.0147 |
| 5 | Minneapolis St. Paul International | 0.0129 |

Hub scores and authority scores are also key measures of the relative importance of a node in a network. In the context of the U.S. domestic airport network, a node with a high hub score represents an airport that offers many flights to many other large and important airports. A high authority score represents an airport that takes in a high volume of traffic from airports with high hub scores. The relationship between hub and authority scores can be seen in Table 2.1 and Table 2.2. It is clear that in the context of a large multi-digraph, such as the U.S. domestic airport network, a high hub score correlates with a high authority score. This intuitive result stems from the structure of our network in which airports with heavy flight traffic service other airports with similar capacity, raising both the hub and authority scores of the airports involved.

**Table 2.1**

| Top 5 Hub Score | | |
|---|---|---|
| Rank | Airport | Score |
| 1 | Chicago O'Hare International | 0.0237 |
| 2 | Hartsfield-Jackson Atlanta International | 0.0201 |
| 3 | Denver International | 0.0196 |
| 4 | Philadelphia International | 0.0178 |
| 5 | Detroit Metro Wayne County | 0.0171 |

**Table 2.2**

| Top 5 Authority Score | | |
|---|---|---|
| Rank | Airport | Score |
| 1 | Chicago O'Hare International | 0.0235 |
| 2 | Hartsfield-Jackson Atlanta International | 0.0209 |
| 3 | Denver International | 0.0193 |
| 4 | Philadelphia International | 0.0184 |
| 5 | Charlotte Douglass International | 0.0172 |

Centrality measures also play a key role in determining the significance of an airport. When developing a list of metrics to determine airport importance, measures of ease of access to other airports and measures of how significant an airport is in connecting other airports are of necessary consideration. Closeness centrality measures the ease of access that one node has to all other nodes in the network. A high closeness centrality score corresponds to greater ease. In the context of our network, high closeness represents airports that have relatively simple flight paths to all other airports. Betweeness centrality measures the significance that a node has in connecting all other nodes in the network. Nodes with high betweeness are critical in connecting nodes throughout the network to each other. Nodes that must be passed through to reach other nodes, or articulation points, generally have high betweeness scores. Table 3 shows the top five airports in our network with the highest closeness centrality scores and betweeness centrality scores. It is interesting to note the emergence of the Ted Stevens Anchorage International Airport in Alaska as an airport with an astoundingly high betweeness centrality and a second place closeness centrality score.

**Table 3**

| Top 5 Closeness | | |
|---|---|---|
| Rank | Airport | Score |
| 1 | Denver International | 0.4848 |
| 2 | Ted Stevens Anchorage International | 0.4838 |
| 3 | Memphis International | 0.4815 |
| 4 | Minneapolis-St Paul International | 0.4813 |
| 5 | Hartsfield-Jackson Atlanta Inter. | 0.4806 |
| Top 5 Betweeness | | |
| Rank | Airport | Score |
| 1 | Ted Stevens Anchorage International | 0.3538 |
| 2 | Seattle/Tacoma International | 0.0873 |
| 3 | Denver International | 0.0757 |
| 4 | Fairbanks International (Alaska) | 0.0684 |
| 5 | Minneapolis-St Paul International | 0.0456 |

The final measure that we took into consideration in determining the most important airports in the U.S domestic airport network was the articulation point property. An articulation point is defined as a node in a network that if removed, causes other nodes to become unreachable. In the context of airport networks, articulation points are of great importance. It is easy to understand that airports which serve as articulation points are vital to the network and pose serious risks to the structure if disrupted. If an airport that is an articulation point is removed, other airports become unreachable, and travelers are unable to reach their specified destinations. Our analysis determined that the U.S domestic airport network contained 92 articulation points, implying that 92 airports served as the only means in which certain airports could reach the rest of the network. To further understand the importance of these articulation points, we explored the question of which articulation points were most significant, specifically which articulation points had the most number of airports depending on their existence. To determine this statistic we developed a simple and effective algorithm.

## 6.1 Articulation Point Importance Algorithm

*1) Calculate # of nodes in Largest Strongly Connected Component $N_1$*
*2) Remove Articulation Point in Question*
*3) Calculate # of nodes in Largest Strongly Connected Component $N_2$*
*4) Calculate Change in # of Nodes: $\Delta_N = N_1 - N_2$*
*5) Iterate through all Articulation Points*

Articulation points with the highest $\Delta_N$ values correspond to the nodes that are the most important articulation points; as removing these nodes cause the most significant decline in the number of nodes in the largest strongly connected component. Applying this algorithm to the 92 articulation points found in the U.S. domestic airport network, we gain insight into which airports are the most significant articulation points. Among some of the airports that serve as the most important articulation points are Guam International, Seattle/Tacoma International, Denver International, and Ted Stevens Anchorage International. Figure 3 shows a visual of the articulation point Ted Stevens Anchorage International Airport (shown as the central yellow node). It is easy to see Ted Stevens Airport's significance as an articulation point. Much of the Alaskan airport network depends on the Ted Steven's Anchorage International Airport to serve as a point to reach the rest of the U.S. network.

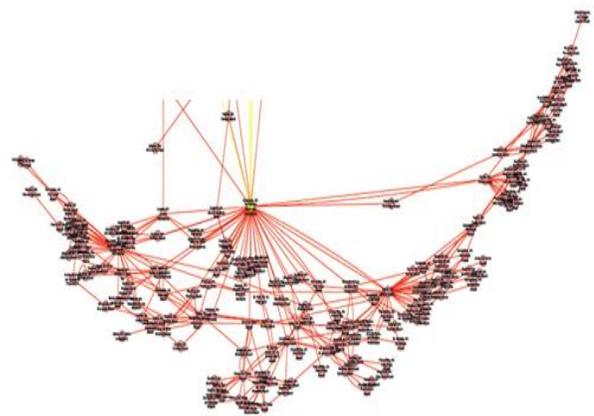

**Figure 3**

## 6.2 Determined Airports

Using the quantitative metrics PageRank, hub score, authority score, centrality, and articulation point significance, we determined the most important airports by qualitatively selecting airports that stood out in all or most measures. Our determination of the most important U.S. domestic airports in regards to their importance in preserving the U.S. domestic airport structure is shown in Table 4 (no particular order).

**Table 4**

| Most Important Domestic Airports (In No Order) |
|---|
| Denver International |
| Ted Stevens Anchorage International |
| Seattle/Tacoma International |
| Minneapolis-St Paul International |
| Hartsfield-Jackson Atlanta International |
| Chicago O'Hare International |

## 7. DISRUPTION IN THE U.S. DOMESTIC AIRPORT NETWORK

After determining which nodes were most important to the U.S. domestic airport network, we examined the effects of disruptions in the form of shut downs to such airports. Disruptions come in a variety of form and magnitude, but for the scope of this paper we only study the disruptions that cause a complete shutdown of the affected airport(s). Real world examples of airport disruptions include natural disaster, terrorist attacks, inclement weather, worker strikes, etc. We study the effects on the network after removing individual important airports and removing collections of important airports. Then we examine the differences between removing our hypothesized important airports and those airports determined by our survey. To begin, we first examine the effects of removing just one airport, specifically the airports from our determined list of important airports. Table 5 shows the changes in network metrics by removing certain airports.

**Table 5**

| Airport | Diameter | | Average Path Length (# trips) | | # of Components | |
|---|---|---|---|---|---|---|
| | *Before* | *After* | *Before* | *After* | *Before* | *After* |
| Denver International | 8 | 8[2] | 3.192 | 3.226 | 1 | 4 |
| Ted Stevens Anchorage International | 8 | 9 | 3.192 | 3.566 | 1 | 4 |
| Chicago O'Hare International | 8 | 8[2] | 3.192 | 3.198 | 1 | 1 |
| Hartsfield-Jackson Atlanta International | 8 | 8[2] | 3.192 | 3.198 | 1 | 2 |
| Minneapolis-St Paul International | 8 | 8[2] | 3.192 | 3.212 | 1 | 2 |
| Seattle/Tacoma International | 8 | 8[2] | 3.192 | 3.219 | 1 | 2 |

Some of the most significant network changes occur when we remove Denver International, Ted Stevens Anchorage International, and Seattle/Tacoma International. Removing Denver International airport effectively breaks up the strongly connected component of U.S. airports into four, separate, unreachable components. The implications of such a disruption are astounding, as complete U.S. travel would no longer be feasible, instead air travel would be reduced to flights within reachable components (four in the case of a Denver International disruption). Average path length provides a good measure of the overall connectedness of the U.S. domestic airport network. It represents the average number of flights needed to take to get between any two airports in the network. Removing Ted Stevens Anchorage International changes the average path length of the U.S. domestic airport network from 3.192 flights to 3.566 flights. On average, flights would require 12% more stops to traverse the network if Ted Stevens Anchorage International were to be out of operation. It is counterintuitive to most to accept this, but upon further inspection we begin to understand why. Ted Stevens Anchorage International serves as a prime airport for both freight flights (those represented in our data with 0 passengers) and passenger flights. Complicated and often lengthy reroute efforts would be required to compensate for the loss of such an airport. The important airports have similar effects on the network upon removal, with average path length and number of strongly connected components increasing across the board. It follows that the average number of flights to get between any two airports increases and certain areas of the U.S. network become unreachable. This resonates the severity of the effects that disruptions to important airports may have.

---

[2] Calculated for LLC because diameter for graph is ∞

We next wanted to examine the resulting network differences between removing our determined important airports and the airports deemed important by our survey. Our survey concluded that the most common airports believed to be vital to the structure of the U.S. domestic airport network were the following: JFK, Los Angeles International, Dulles International (D.C.), Dallas/Fortworth International, and Chicago O'Hare. Using the same definition of a disruption as before (complete shutdown), we simulated a large scale airport disruption on a collection of the important airports determined by us and compared this to a disruption on the set of airports determined by our external survey. As shown in Table 8, we compare complete airport shut downs between set A (our determined top five important airports) and set B (survey result top five important airports). Table 7 provides a key for which airports appear in each set.

**Table 7**

| Set A | Set B |
|---|---|
| Denver International | JFK |
| Ted Stevens Anchorage International | Dulles (Washington D.C.) |
| Chicago O'Hare International | Los Angeles International |
| Hartsfield-Jackson Atlanta International | Dallas/Fortworth International |
| Minneapolis-St Paul International | Chicago O'Hare |

**Table 8**

| Metrics | Full Network | Removing Set B | Removing Set A |
|---|---|---|---|
| Airport | Before | After | After |
| # of Components | 1 | 3 | 9 |
| Average Path Length | 3.192 | 3.224 | 3.632 |
| Diameter | 8 | 8 | 9 |
| # of Articulation Points | 92 | 93 | 95 |

We look to conclude for which set the removal causes more disruption to the U.S. domestic airport network structure. Removing set B causes the U.S. domestic airport to become divided into three strongly connected components, while removing set A results in a network divided into nine strongly connected components. This statistic highlights the greater significance that the airports in set A have in regard to maintaining the structure of the U.S. domestic airport network. Investigating the network consequences further, we see that the average path length of the U.S. domestic airport network after removing the airports in set B is 3.224 flights, while the average path length of the U.S. domestic airport network after removing the airports in Set A is 3.632 flights. Removing the airports of set A causes a 13% larger average path length than does removing the airports of set B. In fact, if the airports of set A were to experience a large scale disruption, the number of flights necessary to travel between airports in the network increases by 14%, while a large scale disruption of the airports in set B only causes the average path length of the network to increase by 1%.

It was also interesting to examine how the removal of set A and set B changed the articulation point dynamics of the U.S. domestic airport network. The removal of set B changed the number of articulation points in the network from 92 to 93, while the removal of set A from the network causes the number of articulation points to increase from 92 to 95. Interpreting these results begins to tell us how removing airports affects the vulnerability of the U.S. domestic airport network. When a network has a higher number of articulation points it becomes inherently more vulnerable to disruptions. This result is intuitive as articulation points are most vital in keeping the network connected. Thus, set A renders itself more significant than set B in preserving the future structure of the U.S. domestic airport network.

Considering the different metrics presented, it is clear that the airports in set A are more important than the airports in set B in maintaining the structure of the U.S. domestic airport network. It is interesting to see that our determined list of important airports contains few airports that most hypothesize to be important.

## 8. DISRUPTION RESPONSE ALGORITHM

Now that we have studied the network consequences of disruptions to our determined list of the most important airports, we look to determine a disruption response method. In particular, we are interested in determining a suitable algorithm that reroutes airport traffic to appropriate airports in the case of a large scale disruption. Based on airport clustering coefficients, distance between airports, and airport capacity we developed a novel algorithm that reroutes airport traffic away from disrupted airports.

The algorithm accepts a theoretically disrupted airport as input and begins by calculating the clustering coefficient of each of its neighboring airports (airports that can be reached within one flight). The clustering coefficient of a node provides a measure of how well connected the neighborhood of the node is. In the context of our airport network, it reveals airports with a similar level of connectivity of the airport in question. Our algorithm then extracts airports with a similar clustering coefficient to our input airport. In our case, clustering coefficients within ±15% of our input airport clustering coefficient were deemed similar. Next, our algorithm iterates through these similar airports, selecting those that fall within a predefined distance from the disrupted airport. In our trial, we set our distance as 200 miles, as further distances become more and more undesirable in rerouting. For all airports that meet the criteria defined above, we calculate the in-degree of each, defined in our network as the number of passengers passing through the airport. This gives us a good estimate of the capacity of the airport which is vital for appropriate rerouting efforts. We also calculate the distance that each airport is from the original disrupted airport. Thus, at this point we have a list of similar airports, within a specified distance from the disrupted airport, and have calculated capacities and distances for each. The final step of our algorithm determines which airports are suitable for rerouting, and the percent of traffic rerouted to each. To calculate this we used the following equation:

*For each airport in our qualifying list: ($x_i$ in X)*

$$\% \text{ Rerouted to } x_i = .85 \times \frac{Capacity(x_i)}{C} + .15 \times \frac{Distance(x_i)}{D}$$

Where:
$D = \sum_{j=1}^{n} Distance(x_j)$ (total sum of distances for all airports in list)
$C = \sum_{j=1}^{n} Capacity(x_j)$ total sum of capacities for all airports in list)

Assigning an 85% weight to capacity and a 15% weight to distance was found to be the "sweet spot" in which reroutings ensured that airports could handle passenger loads while still being convenient to the passengers.

When running our algorithm on a potential disruption of Chicago O'Hare International, our algorithm reroutes 89% of traffic to General Mitchell International in Milwaukee, WI and 11% of traffic to Dubuque Regional Airport, IA. These results are reasonable as the majority of the traffic is rerouted to an airport of similar size 80 miles away, and a small portion of the traffic is routed to a smaller airport located 175 miles away.

Below demonstrates a snip of our code rerouting air traffic from John F. Kennedy International Airport in New York, NY.

>>
*REDIRECTING FLIGHTS FROM: NEW YORK, NY: JOHN F. KENNEDY INTERNATIONAL*
*Philadelphia, PA: Philadelphia International - Distance of 94.0 mi, % Rerouted: 53.585*
*Newark, NJ: Newark Liberty International - Distance of 21.0 mi, % Rerouted: 19.335*
*Hartford, CT: Bradley International - Distance of 106.0 mi, % Rerouted: 9.396*
*Albany, NY: Albany International - Distance of 145.0 mi, % Rerouted 9.205*
*New York, NY: LaGuardia - Distance of 11.0 mi, % Rerouted 6.386*
*Fort Dix, NJ: McGuire Field - Distance of 61.0 mi, % Rerouted: 2.09*

Note that the 54% and 19% of traffic from JFK gets rerouted to Philadelphia International and Newark Liberty International respectively. The remaining traffic is divided up among smaller local airports.

## 9. CONCLUSION

Using network analysis methods on the U.S. domestic airport network is critical in exposing structural holes, determining vital airports, and developing efficient reroutes. By examining key network metrics like PageRank, hub scores, authority scores, centrality measures, and articulation points, we were able to determine that Denver International, Ted Stevens Anchorage International, Seattle/Tacoma International, Minneapolis-St Paul International, Hartsfield-Jackson Atlanta International, and Chicago O'Hare International were the most important airports in maintaining the structure of the U.S. domestic airport network. Information on which airports are most vital to the network is very valuable to both government and commercial agencies. Allocation of funds towards security and maintenance of vital airports can help ensure such airports remain operational.

In addition to determining which airports are most important, our work explores an algorithm used for flight rerouting. Such an algorithm could be of use to flight controllers worldwide. In the case of an airport emergency or disruption, such a tool would provide an efficient way to reroute traffic temporarily.

It was interesting to uncover that expected metrics such as location, size, and number of passengers, did not govern which airports were most important, instead the aforementioned network statistics were the deciding metrics.

Related future works could include the analysis of other major transportation networks, specifically highways, rail travel, and nautical shipping routes. Similar to our work, the analysis of the international flight network would be of interest to government and commercial agencies worldwide. A further breakdown of data by carrier/airline could examine individual company efficiencies.